\documentclass[11pt]{article}
\usepackage[utf8]{inputenc}
\usepackage{calligra}
\usepackage[T1]{fontenc}
\usepackage{titling}
\usepackage{tabularx}
\usepackage{wrapfig}
\usepackage{lipsum}
\usepackage{amsmath,amssymb}
\usepackage{empheq}
\usepackage[linktocpage,
colorlinks=true,
linkcolor=blue,
filecolor=red,      
urlcolor=red,
citecolor=red]{hyperref}
\usepackage{babel}
\usepackage{authblk}
\usepackage[margin = 1 in]{geometry}
\usepackage[dvipsnames]{xcolor}
\usepackage{graphicx}
\usepackage{float}
\usepackage{indentfirst}
\usepackage[labelfont=bf]{caption}
\usepackage{subcaption}
\usepackage{wrapfig}

\newcommand{\mt}{\mathrm}

\numberwithin{equation}{section}
\title{\textbf{Negative Mass Black Holes in de-Sitter Space}}
\author[1,3]{Brayden Hull \thanks{b2hull@uwaterloo.ca}} 
\author[2,3]{Robert B. Mann 
\thanks{rbmann@uwaterloo.ca}}
\affil[1]{\textit{Department of Applied Mathematics, University of Waterloo, Waterloo, Ontario, Canada} }
\affil[2]{\textit{Department of Physics and Astronomy, University of Waterloo, Waterloo, Ontario, Canada} }
\affil[3]{\textit{Perimeter Institute for Theoretical Physics, Waterloo, Ontario, Canada, N2L 2Y5}}
\date{\today}
\begin{document}
\begin{titlingpage}
    \maketitle
    \hrule
    \begin{abstract}
         We show that asymptotically de Sitter black holes of negative mass can exist in Lovelock gravity.  Such black holes have horizon geometries with non-constant curvature and are known as 
         \textit{Exotic Black Holes}. We explicitly examine the case of Gauss-Bonnet gravity.  We briefly discuss the positive mass case where we show how   the  transverse space geometry affects whether a black hole will exist or not. 
         For negative mass solutions we shown how three different black hole spacetimes are possible depending on the transverse space geometry. We also provide closed form expressions for the geometric parameters to ensure that a black hole spacetime is observed. We close with a discussion of the massless case, where there are many different spacetimes that are permitted.
    \end{abstract}
    
  \hrule 
\tableofcontents
\vspace{0.2 in}
\hrule

\end{titlingpage}

\newpage

\section{Introduction}

In General Relativity, the simplest black hole solution is that provided by the Schwarzschild metric \cite{schwarzschild1916}, where the only parameter of the solution is the mass $m$. Less then half a century later a more  general  solution was obtained \cite{KerrNewman}, which describes a black hole with three parameters: mass, angular momentum, and charge.  The introduction of a cosmological constant admits further solutions, as do other matter sources.

In 4 dimensions the assumption that the metric has radial symmetry restricts black hole solutions to have horizons of constant curvature in Einstein gravity. These can be spherical, planar, or hyperbolic  if a negative cosmological constant $\Lambda < 0$ is introduce
\cite{Aminneborg:1996iz,Mann:1996gj,Mann:1997iz}.  If the geometry is hyperbolic then these black holes can have negative mass \cite{Mann_1997} and can form from gravitational collapse \cite{Smith:1997wx}. 

However if $\Lambda \geq 0$ then the situation is notably different. While it has recently been shown that de Sitter Schwarzschild spacetimes with negative mass in \cite{johnson2020stable,mbarek2014negative} can exist in Einstein gravity,
these solutions consist of  spacetime bubbles. No negative mass black hole vacuum solutions have been obtained.

 We show in this paper that 
 asymptotically de Sitter   black hole vacuum solutions of negative mass   exist in   Lovelock gravity.  This somewhat unexpected finding arises from a feature recently pointed out for black holes in higher-curvature gravity: their event horizons   need not be of constant curvature \cite{Ray_2015,Dotti_2005,Dotti_2007,DOTTI_2009,Dotti_2010,Oliva_2013}. The transverse space of the (potential) black hole solution need only be  an Einstein space, admitting horizon  geometries of increasing complexity as the spacetime dimension increases. For this reason these have been referred to as \textit{Exotic Black Holes} (EBHs). 
 
 We demonstrate that static asymptotically de Sitter Lovelock EBHs can have negative mass. We explicitly illustrate this for Gauss-Bonnet gravity, but it is clear that this will occur in higher-order Lovelock gravity and we expect in more general higher-curvature theories as well.
 Lovelock gravity is a particular subclass of  higher curvature gravity theories, which  have garnered much attention since quantum gravity induces such higher-order corrections  to the standard
 Einstein-Hilbert gravitational action \cite{birrell_davies_1982}. Furthermore, renormalization properties are improved for such theories  \cite{StelleR}.
 The Lovelock class \cite{Lovelock2} is of particular interest: it is regarded as the most natural higher-curvature generalization of Einstein gravity whose field equations are of  second order in the metric functions.

Research into EBHs has been of recent interest  particularly in the thermodynamics of asymptotically Anti de Sitter black holes. The geometry of event horizon was shown to greatly affect the thermodynamic phenomena \cite{Hull2021,HullSimovic,Farhangkhah:2014zka,Farhangkhah:2021tzq}. While investigation of black holes in AdS space has been of great interest since the discovery the AdS/Conformal Field Theory correspondence  \cite{Witten:1998zw,Maldacena}, interest in de Sitter black holes is motivated primarily by cosmological considerations \cite{planck2016-l06}, though some motivation comes from considerations of holographic duality  \cite{strominger2001ds}.

We begin with a brief review of Lovelock gravity and EBHs before moving onto Gauss-Bonnet solutions. We provide general polynomial equations for determining horizon locations as well as the Kretschmann scalar, which will function as a diagnostic for finding spacetime singularties.   We then discuss the positive mass case and how horizon geometry  can dictate whether or not a black hole may exist. We then provide an  analysis of the negative mass case, where we obtain a bound on the minimum allowed mass for obtaining a black hole. We also present a closed form expression that  dictates the bounds on the horizon geometry for which   a negative mass black hole is admitted as a vacuum solution.   We have   shown that it is possible to obtain massless de Sitter black holes with the proper set of parameters, and discuss this in the penultimate section before summarizing our work.

\section{Lovelock Gravity \& \textit{Exotic Black Holes}}

For a Lovelock theory of gravity the Lagrangian density is given by
\begin{equation} \label{Lag}
    \mathcal{L}=\frac{1}{16 \pi G_{N}} \sum_{k=0}^{K} \hat{\alpha}_{k} \mathcal{L}^{(k)}
\end{equation}
where $\hat{\alpha}_{k}$ are the Lovelock coupling constants and $\mathcal{L}^{(k)}$ are the Euler densities with dimension $2k$, which are  
\begin{equation}
    \mathcal{L}^{(k)}=\frac{1}{2^{k}} \delta_{c_{1} d_{1} \ldots c_{k} d_{k}}^{a_{1} b_{1} \ldots a_{k} b_{k}} R_{a_{1} b_{1}}^{c_{1} d_{1}} \ldots R_{a_{k} b_{k}}^{c_{k} d_{k}}
\end{equation}
with $\delta$ representing the generalized fully anti-symmetric Kronecker delta. The first few terms are
$$
\mathcal{L}^{(0)}=1 \qquad \mathcal{L}^{(1)}=R  \qquad \mathcal{L}^{(2)}=R^{2}-4 R_{a b } R^{a b }+R_{a b c d} R^{a b c d}
$$ 
with $R$ the Ricci Scalar and $\mathcal{L}^{(2)}$
 the Gauss-Bonnet term. From the Lagrangian density \eqref{Lag} we may construct an action 
\begin{equation}
    S=\int d^{d}x \sqrt{-g}\left( \frac{1}{16 \pi G_{N}} \sum_{k=0}^{K} \alpha_{k} \mathcal{L}^{(k)}+\mathcal{L}_{m}\right) 
\end{equation}
 including a matter term $\mathcal{L}_{m}$.
Variation with respect to the metric $g^{ab}$ yields the field equations
\begin{equation} \label{fieldeq}
\sum_{k=0}^{K} \hat{\alpha}_{(k)} \mathcal{G}_{a b}^{(k)}=8 \pi G_{N} T_{ a b}.
\end{equation}
where $T_{ab}$ is the stress energy tensor of the matter field and $\mathcal{G}_{ab}^{(k)}$ are the Lovelock tensors 
\begin{equation}
    \mathcal{G}_{a b}^{(k)}=-\frac{1}{2^{(k+1)}} g_{a b} \delta_{ e_{1} f_{1} \ldots e_{k} f_{k}}^{ c_{1} d_{1} \ldots c_{k} d_{k}} R_{c_{1} d_{1}}^{e_{1} f_{1}} \ldots R_{c_{k} d_{k}}^{e_{k} f_{k}}
\end{equation}

We will consider only vacuum solutions with $\mathcal{L}_{m}=0$ henceforth, so the field equations  will be \eqref{fieldeq} with the stress energy being set to zero. The parameter $K \leq \frac{d-1}{2}$, where $d$ is the dimension of spacetime, and sets the maximal degree of non-linearity in the curvature.  Values of $K$ larger than this make no contribution to the field equations and are topological invariants in the action.

The metric ansatz we will make use of is 
\begin{equation} \label{metric}
\begin{aligned}
d s^{2} &=\textsf{g}_{i j} d y^{i} d y^{j}+\gamma_{\alpha \beta} d x^{\alpha} d x^{\beta} \\
&=-f(r) d t^{2}+\frac{d r^{2}}{f(r)}+r^{2} d \Sigma_{d-2}^{2}
\end{aligned}
\end{equation}
where $y^{i}=(t,r)$ and $\textsf{g}_{ij}=\mt{diag}(-f(r),\frac{1}{f(r)})$. The  line element   $ d\Sigma^{2}_{d-2}$ of the base manifold (or transverse space) 
has metric $\gamma_{\alpha\beta}$
and coordinates $x^\alpha$. We shall assume it to be compact, with volume
$\Sigma_{d-2}$, the analogue of the volume of a unit sphere.  If this space is not compact then quantities
such as $M/\Sigma_{d-2}$
will be regarded as finite.

The only other condition on the transverse space is that it is an Einstein space, $R_{\alpha\beta}\propto \gamma_{\alpha\beta}$. The most commonly studied special case is when the base space has constant curvature, with curvature parameter $\kappa=(-1,0,1)$, which corresponds to negative, flat and positive curvature respectively. Solutions in which this latter condition is relaxed, so that the transverse space is only an Einstein space \cite{Dotti_2005}, shall be referred to as  \textit{Exotic Black Holes} \cite{Ray_2015,RayExBH}. 

Introducing this metric into the field equations yields a resulting polynomial equation for $f(r)$ \cite{Ray_2015}
\begin{equation} \label{poly}
    \sum_{n=0}^{K} \frac{b_{n}}{r^{2 n}}\left(\sum_{k=n}^{K} \alpha_{k}\binom{k}{n}\left(\frac{-f(r)}{r^{2}}\right)^{k-n}\right)=\frac{16 \pi G_{N} M}{(d-2) \Sigma_{d-2} r^{d-1}}
\end{equation}
where $M$ represents the mass of the  black hole and $\Sigma_{d-2}$  the volume of the base space. The quantities $\alpha_{k}$ are rescaled Lovelock couplings, defined as
\begin{equation}
    \alpha_{0}=\frac{\hat{\alpha}_{0}}{(d-1)(d-2)}, \quad \alpha_{1}=\hat{\alpha}_{1}, \quad \alpha_{k}=\hat{\alpha}_{k} \prod_{n=1}^{2 k}(d-n) \quad \text { for } k \geq 2.
\end{equation}
The cosmological constant is defined through the zeroth Lovelock coupling constant $\Lambda=-\frac{\hat{\alpha}_{0}}{2}$. 

Imposing only the condition that the transverse space is an Einstein space admits the following possibilities
\begin{equation}\label{exoticeqs}
    \begin{aligned}
    \hat{\mathcal{G}}_{\beta}^{(n) \alpha} &=-\frac{(d-3) ! b_{n}}{2(d-2 n-3) !} \delta_{\beta}^{\alpha}\\
\hat{\mathcal{L}}^{(n)} &=\frac{(d-2) ! b_{n}}{(d-2 n-2) !} 
\end{aligned}
\end{equation}
for its Lovelock tensor and associated intrinsic Euler density respectively.
The constants  $b_{n}$  we refer to as  \textit{topological terms}, and can take any value in $\mathbb{R}$.  They define the topology of the base manifold and the geometry of the event horizon from the condition $f(r_{+})=0$ that defines the horizon location $r_+$. Without  loss of generality we  set $b_{0}=1$;  to recover Einstein gravity in the proper limit of vanishing $\alpha_{k \geq 2}$,
we will set $\alpha_{1}=1$. The mass $M$ is defined through the Hamiltonian formalism, and is the conserved charge of the timelike Killing vector of the background spacetime. When the dimension is  $d=2K+1$ the transverse space is restricted to be of constant curvature \cite{Dotti:2010bw,Dotti:2007az}.

The Kretschmann scalar for a metric  of the form  \eqref{metric} is 
\begin{equation}\label{Kret}
    R^{a b c d} R_{a b c d}=\left(\frac{d^{2} f(r)}{d r^{2}}\right)^{2}+2 \frac{(d-2)}{r^{2}}\left(\frac{d f(r)}{d r}\right)^{2}+2 \frac{(d-2)(d-3) f(r)^{2}}{r^{4}}-4 \frac{R[\gamma] f(r)}{r^{4}}+\frac{\mathcal{K}[\gamma]}{r^{4}}
\end{equation}
where $R[\gamma]$ and $\mathcal{K}[\gamma]$ are the Ricci  and Kretschmann scalars of the tranverse space respectively.

\section{Gauss-Bonnet Solutions}

Setting $K=2$ in   \eqref{poly} yields
\begin{equation}
    \frac{\alpha_{2} f^{2}}{r^{4}}+\left(-\frac{1}{r^{2}}-\frac{2 b_{1} \alpha_{2}}{r^{4}}\right) f+\alpha_{0}+\frac{b_{1}}{r^{2}}+\frac{b_{2} \alpha_{2}}{r^{4}}=\frac{16 \pi M}{(d-2) \Sigma_{d-2} r^{d-1}}
\end{equation}
with the solutions
\begin{equation}
    f=f_{\pm}(\mathrm{m}) \equiv \frac{r^{2}+2 b_{1} \alpha_{2} \pm \sqrt{\left(b_{1}^{2}-b_{2}\right) 4 \alpha_{2}^{2}+r^{4}\left(1-4 \alpha_{2} \alpha_{0}\right)+\frac{8\mathrm{m} \alpha_{2}}{r^{d-5}}}}{2 \alpha_{2}}
\end{equation}
where we have written
\begin{equation}
    \mathrm{m} \equiv \frac{8 \pi M}{(d-2) \Sigma_{d-2}}.
\end{equation}
The two solutions are distinguished by their behaviour in the $\alpha_{2}\rightarrow 0$ limit. The $f_{-}$ branch is referred to as the Einstein branch as it has a smooth $\alpha_{2}\rightarrow 0$ limit, giving  
\begin{equation}
    \lim _{\alpha_{2} \rightarrow 0} f_{-}(\mathrm{m})=\alpha_{0} r^{2}+b_{1}-\frac{2 \mathrm{m}}{r^{d-3}}
\end{equation}
whereas the $f_{+}$ solution  does not have a smooth limit as $\alpha_{2}\rightarrow0$; it is referred to as the Gauss-Bonnet branch.

For the remainder of this paper we will only consider the Einstein branch for analysis; we will also choose $b_{1}=1$, so that as $\alpha_{2}\rightarrow 0$  we recover the standard Schwarzschild
(Anti) de Sitter 
solution. We will only examine  solutions with $d \geq 6$, since the $d=5$ case only admits  constant curvature solutions
in the transverse space.  The cosmological constant is 
\begin{equation}
    \begin{aligned}
    \Lambda&=-\frac{\hat{\alpha}_{0}}{2} 
=-\frac{\alpha_{0}(d-1)(d-2)}{2}
    \end{aligned}
\end{equation}
and we shall set $\alpha_{0}<0$ since we are interested in  asymptotically de Sitter solutions.  This allows us to write 
\begin{align}
    f_{-}&=\frac{r^{2}+2  \alpha_{2} -\sqrt{\left(1-b_{2}\right) 4 \alpha_{2}^{2}+r^{4}\left(1+4 \alpha_{2} \frac{2\Lambda}{(d-1)(d-2)}\right)+\frac{8\mathrm{m} \alpha_{2}}{r^{d-5}}}}{2 \alpha_{2}}\\
\end{align}
where we note $\Lambda>0$.

It is  convenient at this point to introduce a set of dimensionless variables 
\begin{equation}
    r=x \sqrt{\alpha_{2}}, \quad \Lambda=\frac{(d-1)(d-2)z}{2\alpha_{2}}, \quad \mathrm{m}=m \alpha_{2}^{\frac{d-3}{2}},\quad z>0
\end{equation}
so that  
\begin{equation} \label{feq}
    f=1+\frac{x^2}{2}-\frac{\sqrt{(1+4 z) x^{4}+4(1-b_{2})+\frac{8 m}{x^{d-5}}}}{2}.
\end{equation}
  The horizon(s) are located at the roots of $f$, which can be found from solving the polynomial:
\begin{equation}\label{horloc}
     x^{d-1} z- x^{d-3}- b_{2} x^{d-5}+2 m=0 \quad \mathrm{with}\quad  d\geq 6, \quad z>0
\end{equation}
For a   black hole solution we must have  
$f(x)>0 \; \mathrm{for} \; x>x_{+}$ for some range of $x$, where $x_{+}$ locates the outermost event horizon of the black hole.

Superfically it appears that there will be 2 or 0  positive roots to this polynomial from Descartes' rule of signs. However if we
 allow both $b_{2}$ and  $m$ to be either positive or negative the number of possible positive roots are then  the following:
\begin{align}
    & \mt{If} \ m>0  &\mathrm{then} \ 2\; \textrm{roots} \rightarrow (x_{c},x_{+}) \ \mathrm{or}\  0\ \textrm{roots for any}\;  b_2 & \nonumber \\
    & \mt{If} \ m<0\ \textrm{and}  \ b_{2}>0,  &\mathrm{then}\ 1\  \mt{root}\rightarrow x_{c} \hfill & \nonumber \\ &\mt{If} \ m<0\ \textrm{and}  \ b_{2}<0,  &\mathrm{then}\ 3\  \mt{roots} \rightarrow (x_{-},x_{+},x_{c})\  \mt{or} \ 1 \  \mt{root}\rightarrow x_{c}. &
\end{align}
where  $x_{c}$ is the cosmological horizon, $x_{-}$ is the inner horizon, and $x_{+}$ is  the outer horizon of the black hole. 

\subsection{Positive Mass}

\begin{figure}[H]
    \centering
    \begin{subfigure}{0.32\textwidth}
    \centering
    \includegraphics[width=\textwidth]{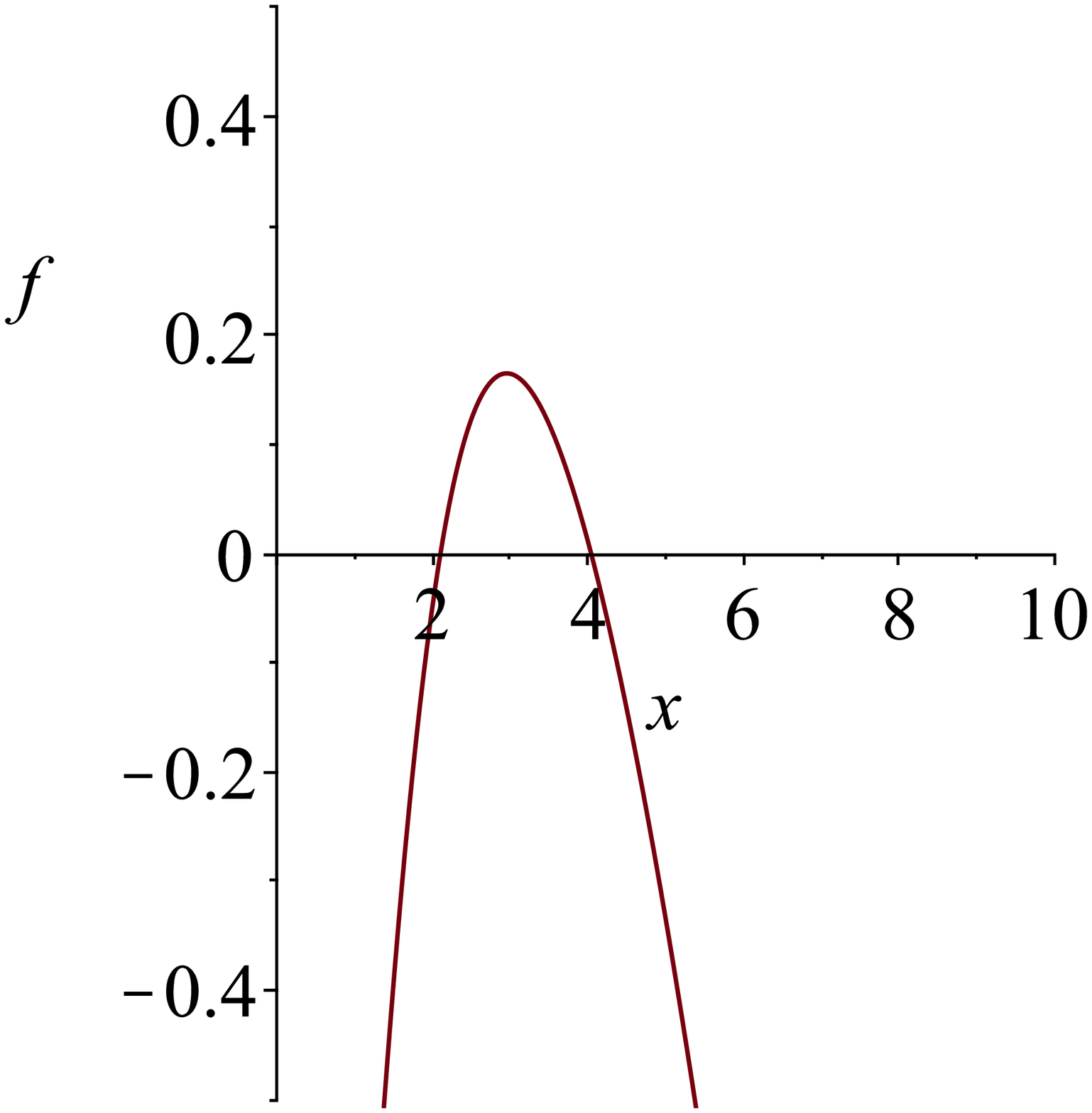}
    \end{subfigure}
    \begin{subfigure}{0.32\textwidth}
    \centering
    \includegraphics[width=\textwidth]{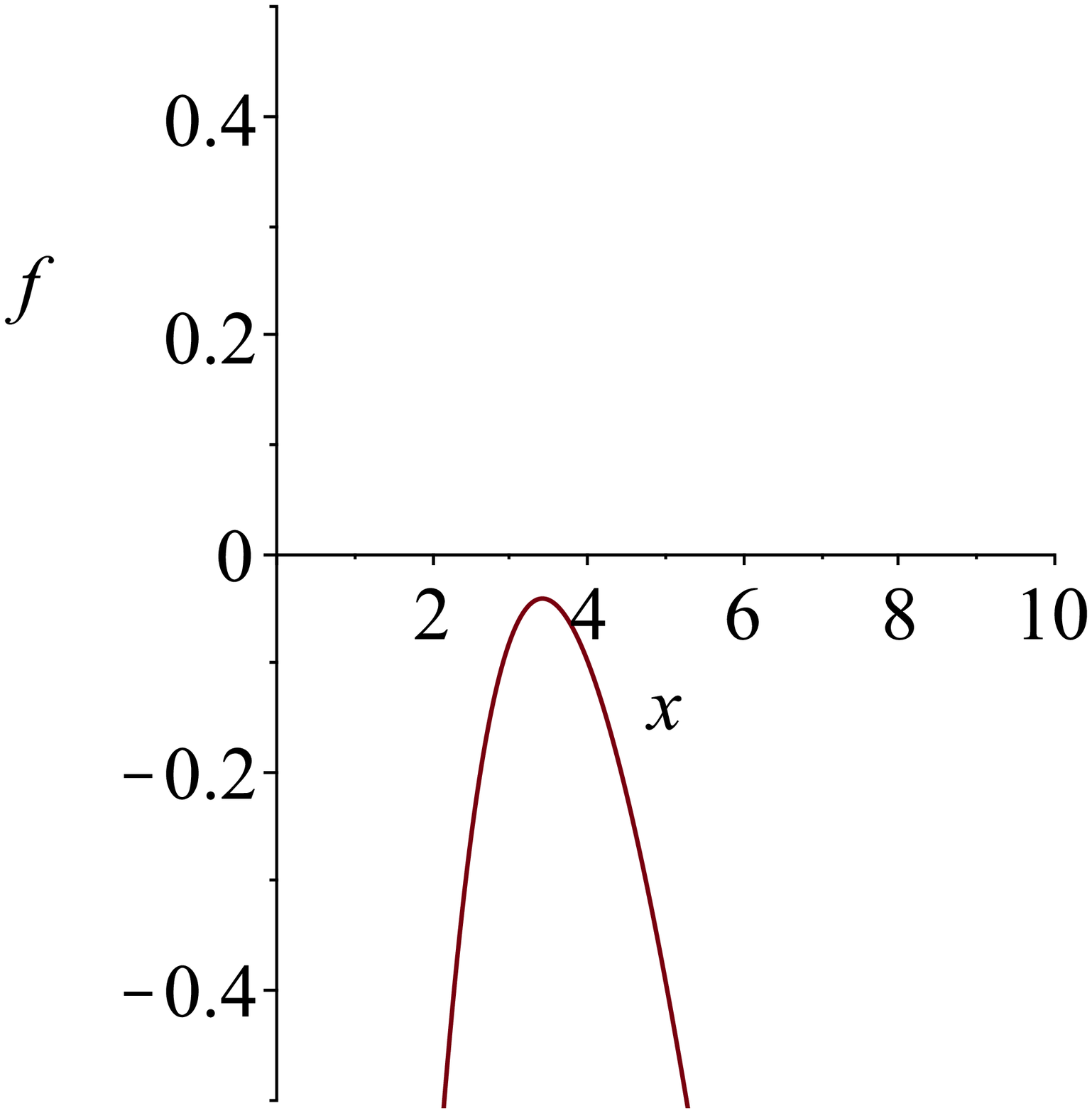}
    \end{subfigure}
    \begin{subfigure}{0.32\textwidth}
    \centering
    \includegraphics[width=\textwidth]{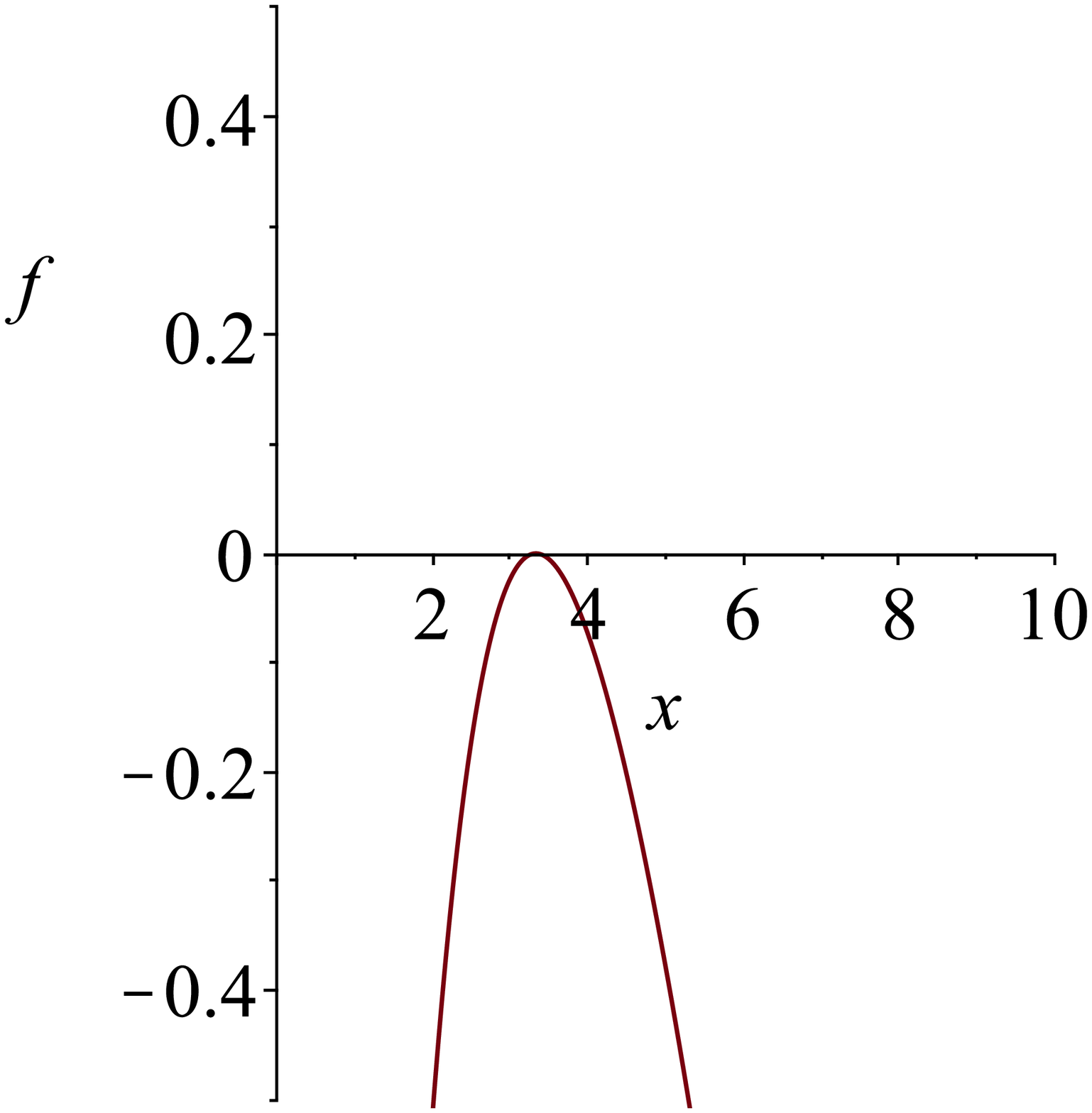}
    \end{subfigure}
    \caption{\textbf{Positive Mass Black Hole}\ $b_{2}=-2.5,z=0.5,d=6$.\ \textbf{Left:} Black hole with mass $m=1$ showing two distinct real roots. \textbf{Center:} Mass $m=5$, showing no roots, and a naked singularity. \textbf{Right:} Mass $m=4.0638$, showing a double root. }
    \label{PositiveMass}
\end{figure}

Consider first positive mass. We plot the behaviour of $f$  in Figure~\ref{PositiveMass} for sample values  $b_{2}=-2.5$, $z=0.05$, and $d=6$,   varying $m$.  For small nonzero $m$ we have an asymptotically de Sitter black hole (left diagram in Figure~\ref{PositiveMass}) where as for sufficiently large $m$ there is no static black hole solution and we have an expanding spacetime with $r$ and $t$ interchanging roles (middle diagram).  As $m$ increases from small to large values, there is one value of $m$ where $f$ has a double root and the horizons merge, illustrated in the right  diagram in Figure~\ref{PositiveMass}. This is a Gauss-Bonnet generalization of the Nariai solution.  This structure is the same in all dimensions $d \geq 6$. Only quantitative differences are seen as we increase the dimension:   the spread between $x_{+}$ and $x_{c}$ increases and the maximum value of $f$   slightly increases. Consequently we will illustrate our results in 6 dimensions unless otherwise stated.

\begin{figure}[H]
    \centering
    \begin{subfigure}{0.4\textwidth}
    \centering
    \includegraphics[width=\textwidth]{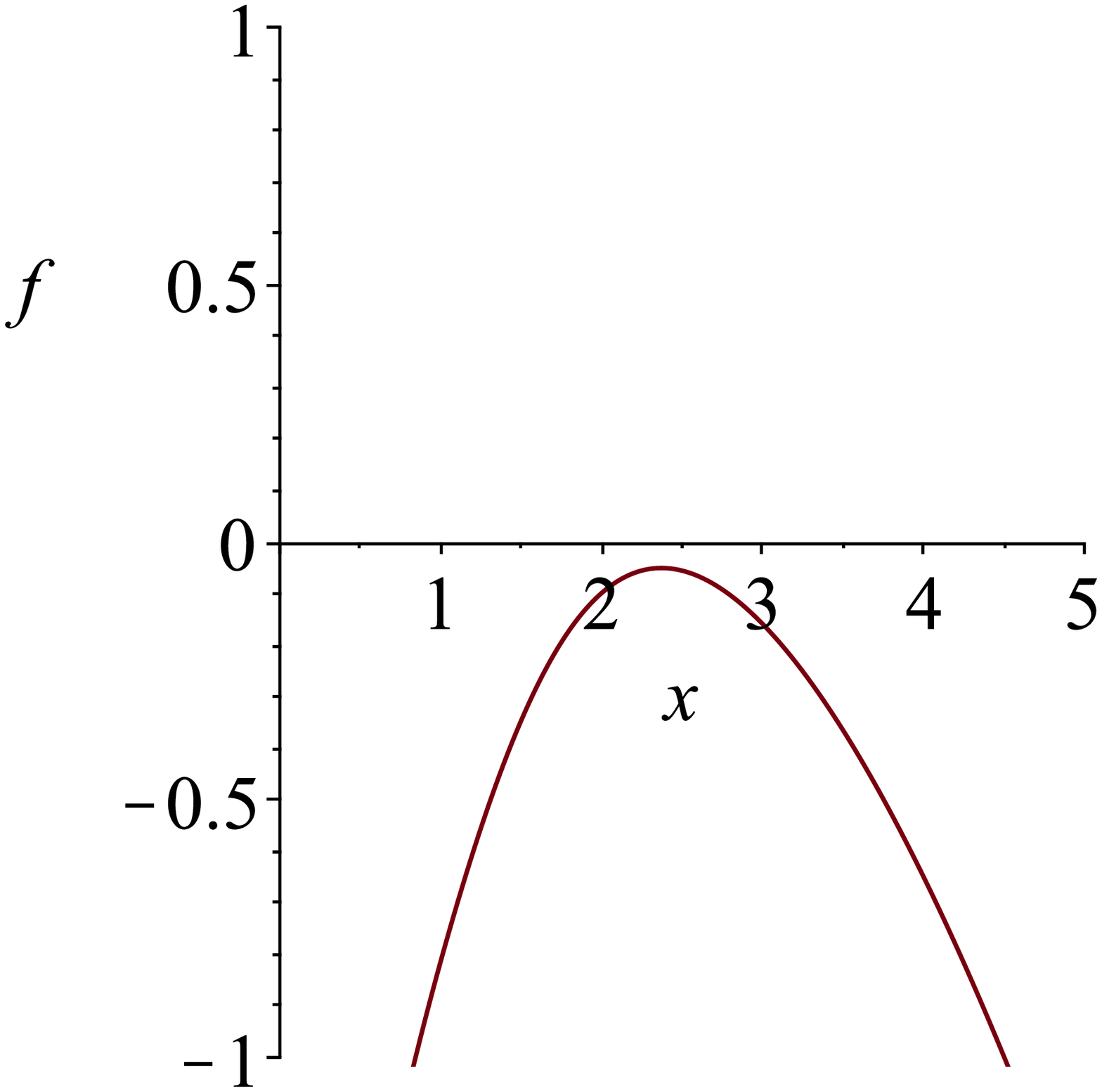}
    \end{subfigure}
    \begin{subfigure}{0.4\textwidth}
    \centering
    \includegraphics[width=\textwidth]{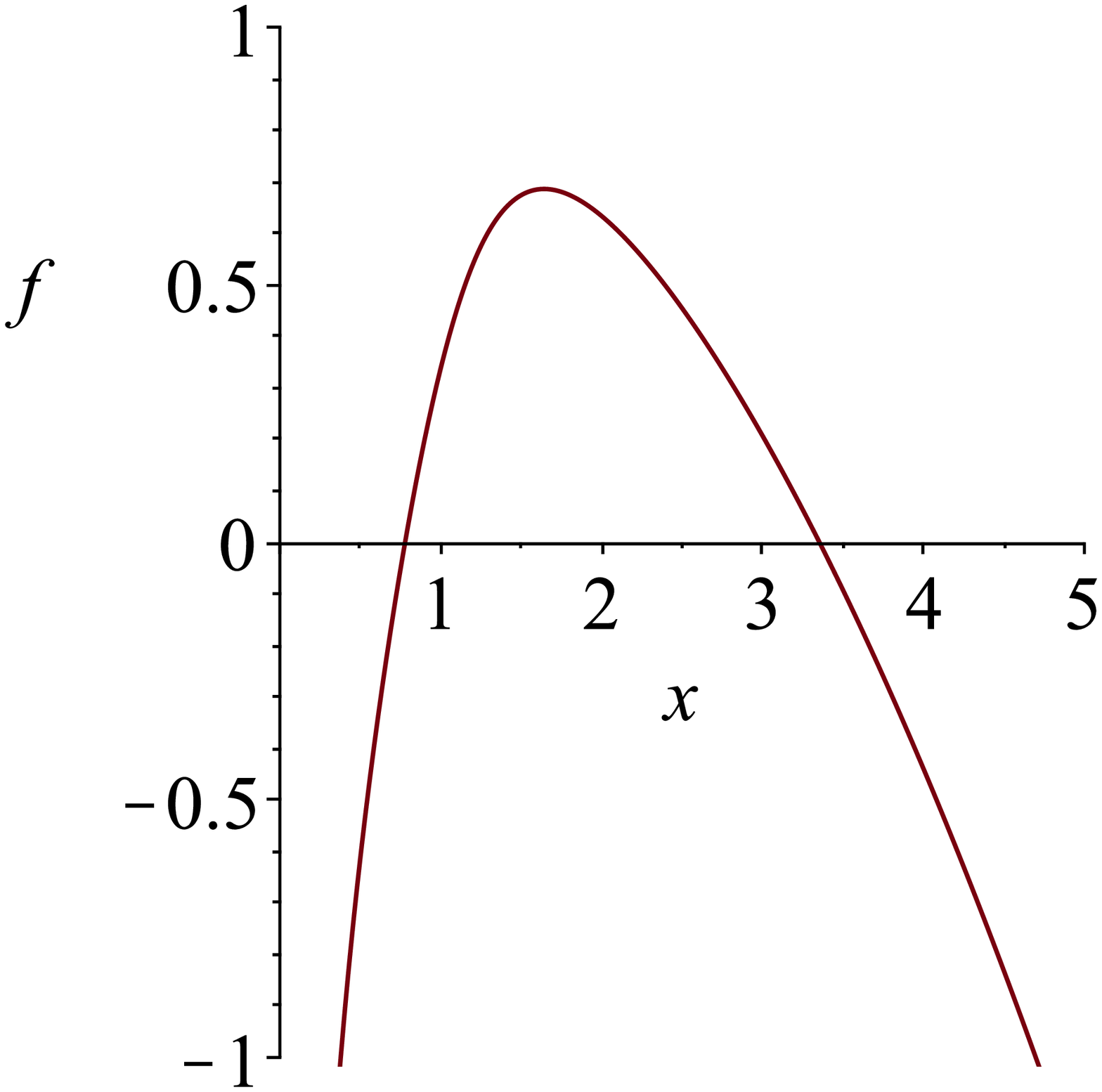}
    \end{subfigure}
    \caption{\textbf{Positive Mass}\ $m=1,z=0.1,d=6$.\ \textbf{Left:} Naked singularity with $b_{2}=-2$ showing no real roots. \textbf{Right:} de Sitter black hole with $b_{2}=2$ show a cosmological horizon and event horizon $x_{c}>x_{+}$}
    \label{diffb2}
\end{figure}

We may also examine what happens when we hold all the parameters fixed except for $b_{2}$, which is seen in Figure~\ref{diffb2}. For a fixed mass $m$ and cosmological parameter $z$ increasing the value of $b_{2}$ will shift $f$ upwards. In this manner the  geometry of the horizon plays an important role regarding the existence of a black hole solution, namely that a minimum allowed value of $b_{2}$ is required.  This minimum value occurs when $f$ has a double root, corresponding to  a   Nariai type of black hole solution.  In de Sitter space this is referred to as the maximum allowed mass  a black hole can have, other parameters being fixed. To determine the value of
$x_+=x_c$ for the Nariai-type solution, the   equations 
\begin{equation}
    f=0,\quad \frac{\partial f}{\partial x}=0
\end{equation}
must be simultaneously solved.
In 6 dimensions there is no closed form solution since the latter equation is  quintic polynomal is $x$. For the parameters  in Figure 2 we obtain numerically $b_{{2}_{min}} \approx -1.62$. For $b_{2}$ less than this  the spacetime  will have a naked singularity     at the origin. 

\begin{figure}[H]
    \centering
    \begin{subfigure}{0.41\textwidth}
    \centering
    \includegraphics[width=\textwidth]{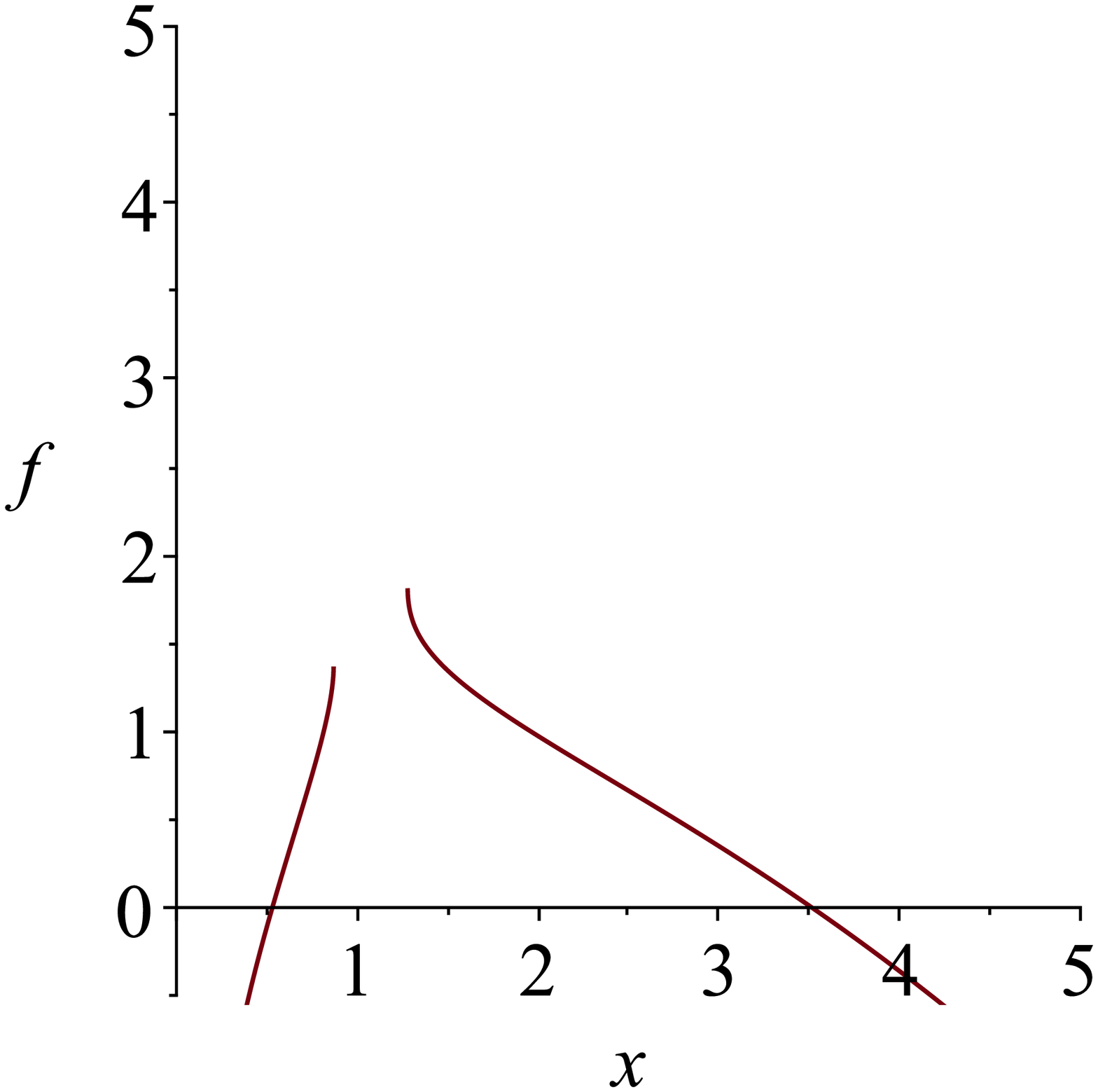}
    \end{subfigure}
    \begin{subfigure}{0.41\textwidth}
    \centering
    \includegraphics[width=\textwidth]{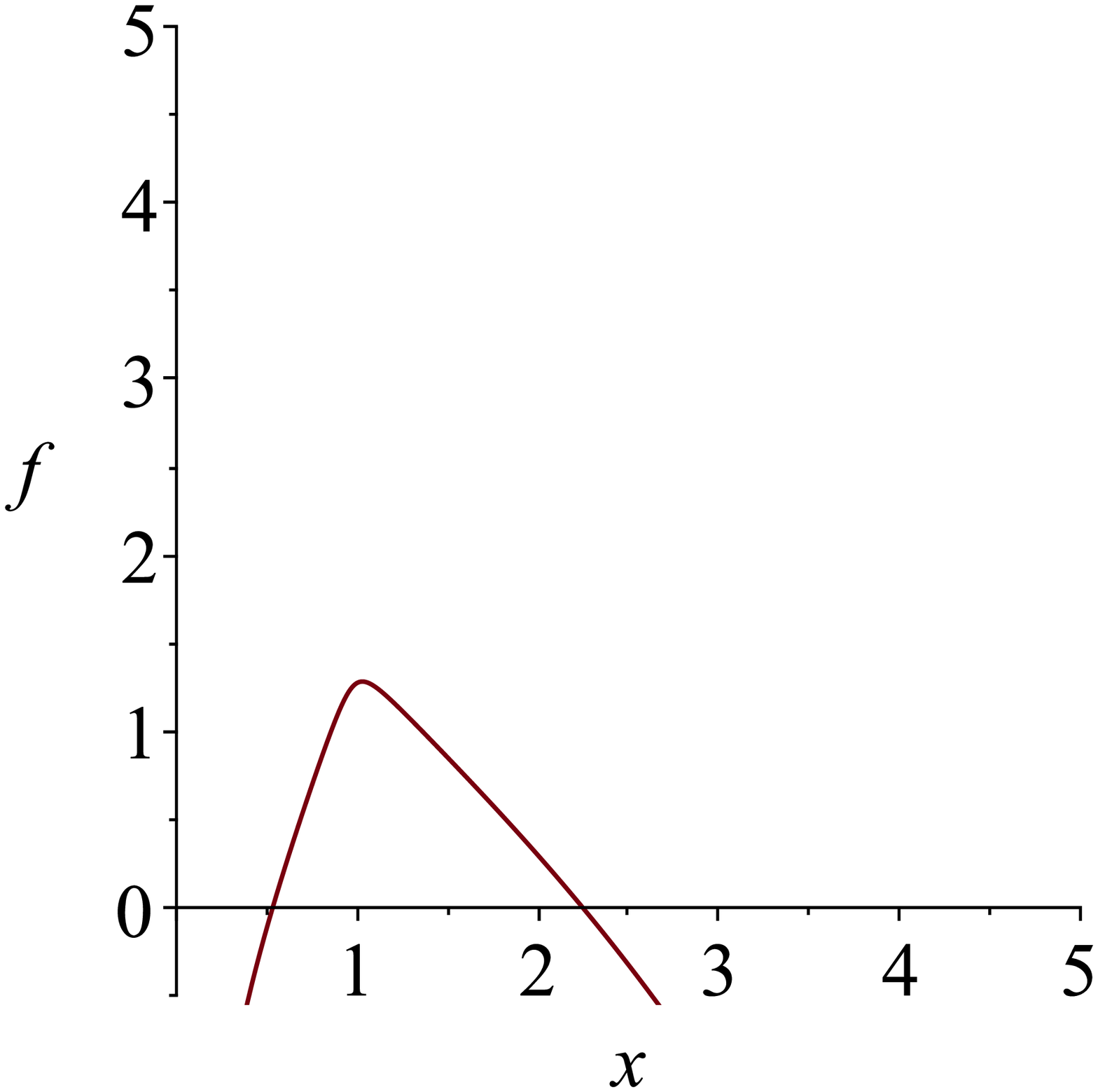}
    \end{subfigure}
    \caption{\textbf{Positve mass } $m=1$,$d=6$,$b_{2}=3.5$. \textbf{Left:}$z=0.1$ Two discontinuous branches with two spacetime singularities. \textbf{Right:} $z=0.3$ de Sitter black hole.}
    \label{fbreak}
\end{figure}

We can also make another observation, shown in Figure~\ref{fbreak}. That is, for a certain set of parameters $f$ can become discontinuous for some range of $x$. This discontinuity is due to the argument inside the square root becoming negative. Whenever this occures the Kretschmann scalar \eqref{Kret} diverges, yielding a spacetime singularity.  Solutions to the following polynomial
\begin{equation} \label{fbeq}
    \left(4-4 b_{2}\right) x^{d-5}+(4 z+1) x^{d-1}+8 m=0
\end{equation}
yield the   singularity location(s). 
From Decartses' rule of signs this will only have a root when $b_{2}>1$. This is a necessary but not sufficient condition; as illustrated  in the right image of Figure~\ref{fbreak}, increasing the value of $z$  allows the two discontinuous branches to  join, yielding   a smooth continuous metric function $f$.  If we hold all parameters fixed, increasing $z$ shifts $f$ down vertically and reduces the spread between the black hole and cosmological horizons, allowing the possibility of a smooth metric function between them.

\subsection{Negative Mass}

\begin{figure}[H]
    \centering
    \begin{subfigure}{0.32\textwidth}
    \centering
    \includegraphics[width=\textwidth]{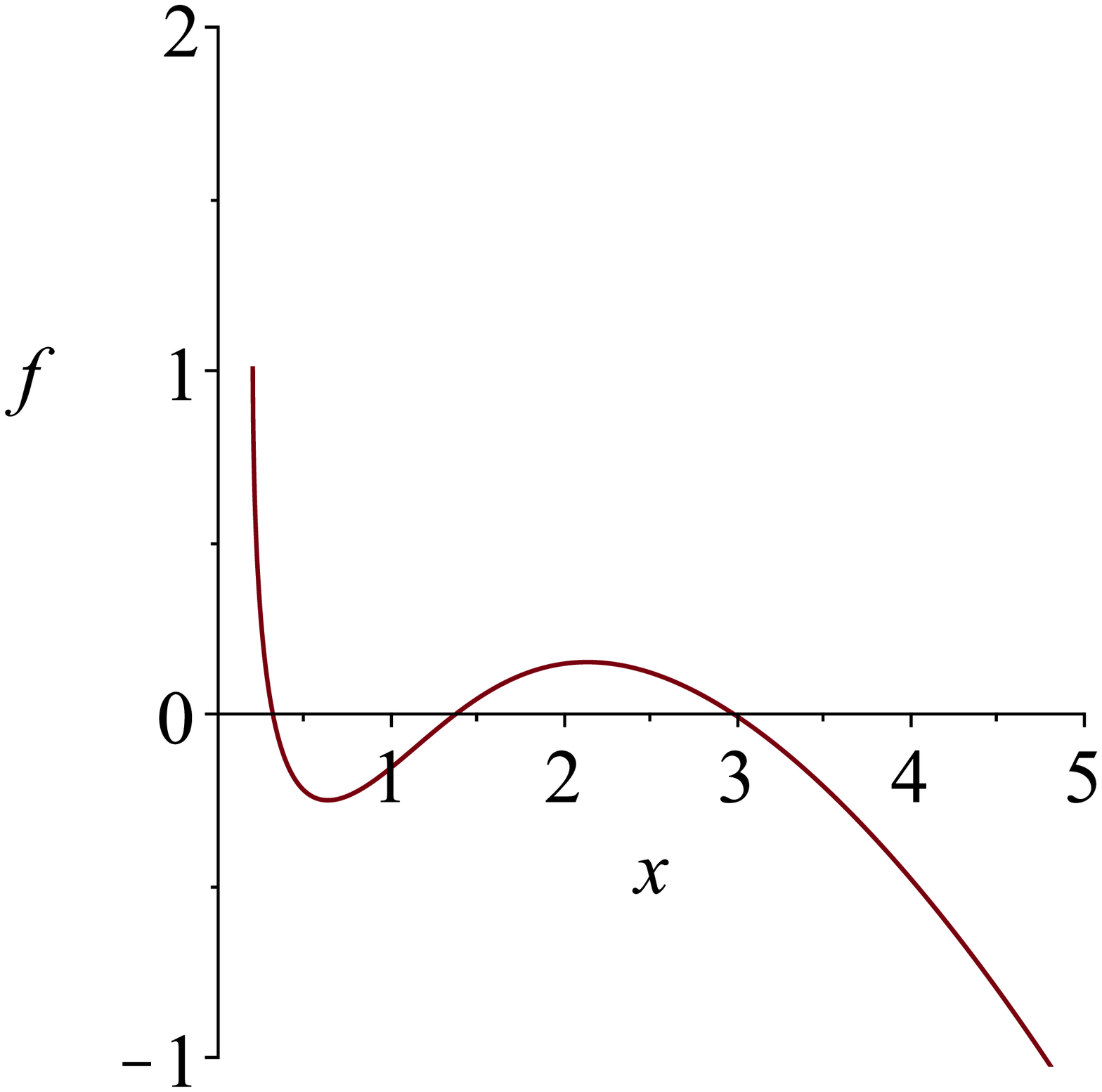}
    \end{subfigure}
    \begin{subfigure}{0.32\textwidth}
    \centering
    \includegraphics[width=\textwidth]{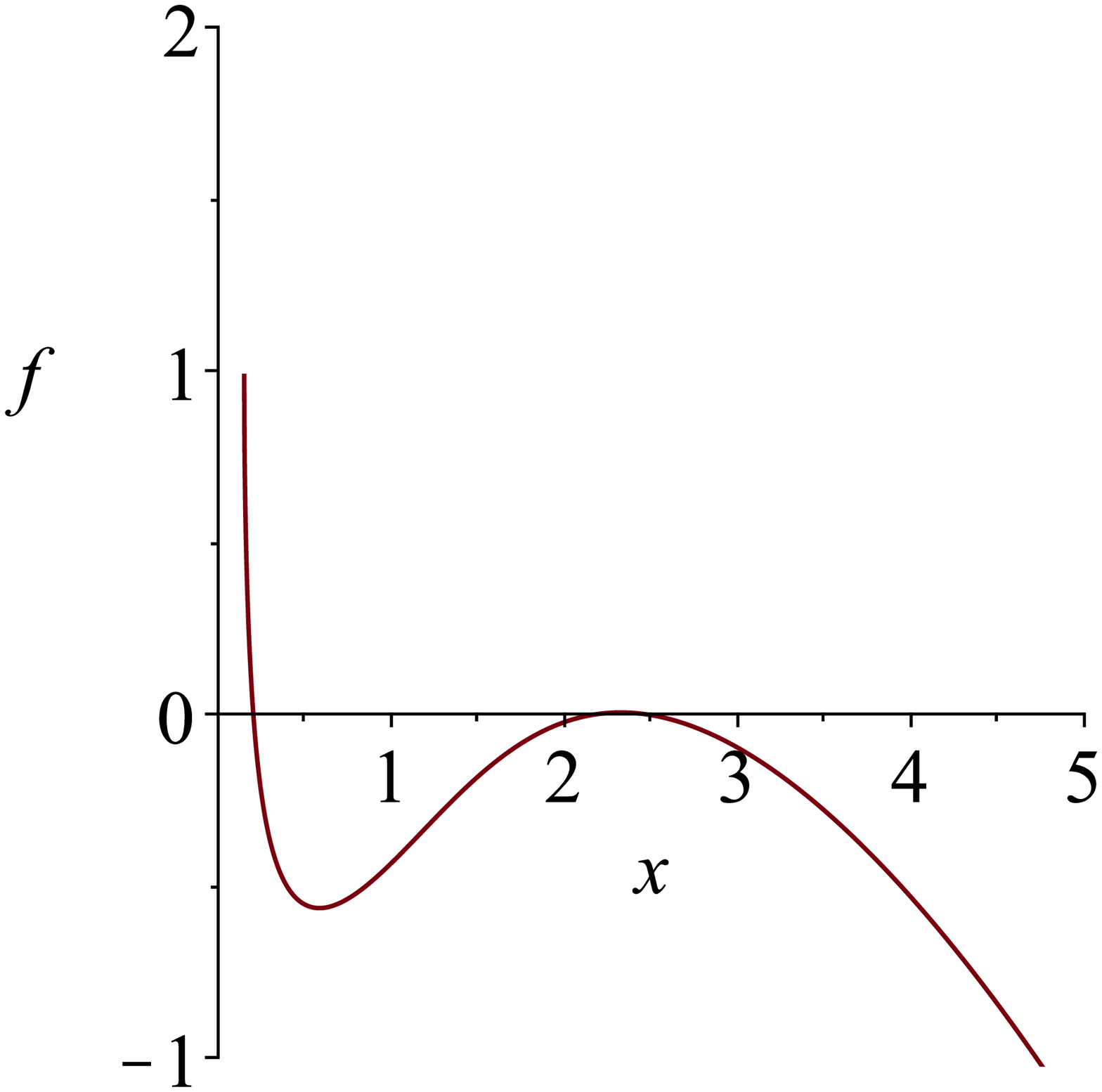}
    \end{subfigure}
    \begin{subfigure}{0.32\textwidth}
    \centering
    \includegraphics[width=\textwidth]{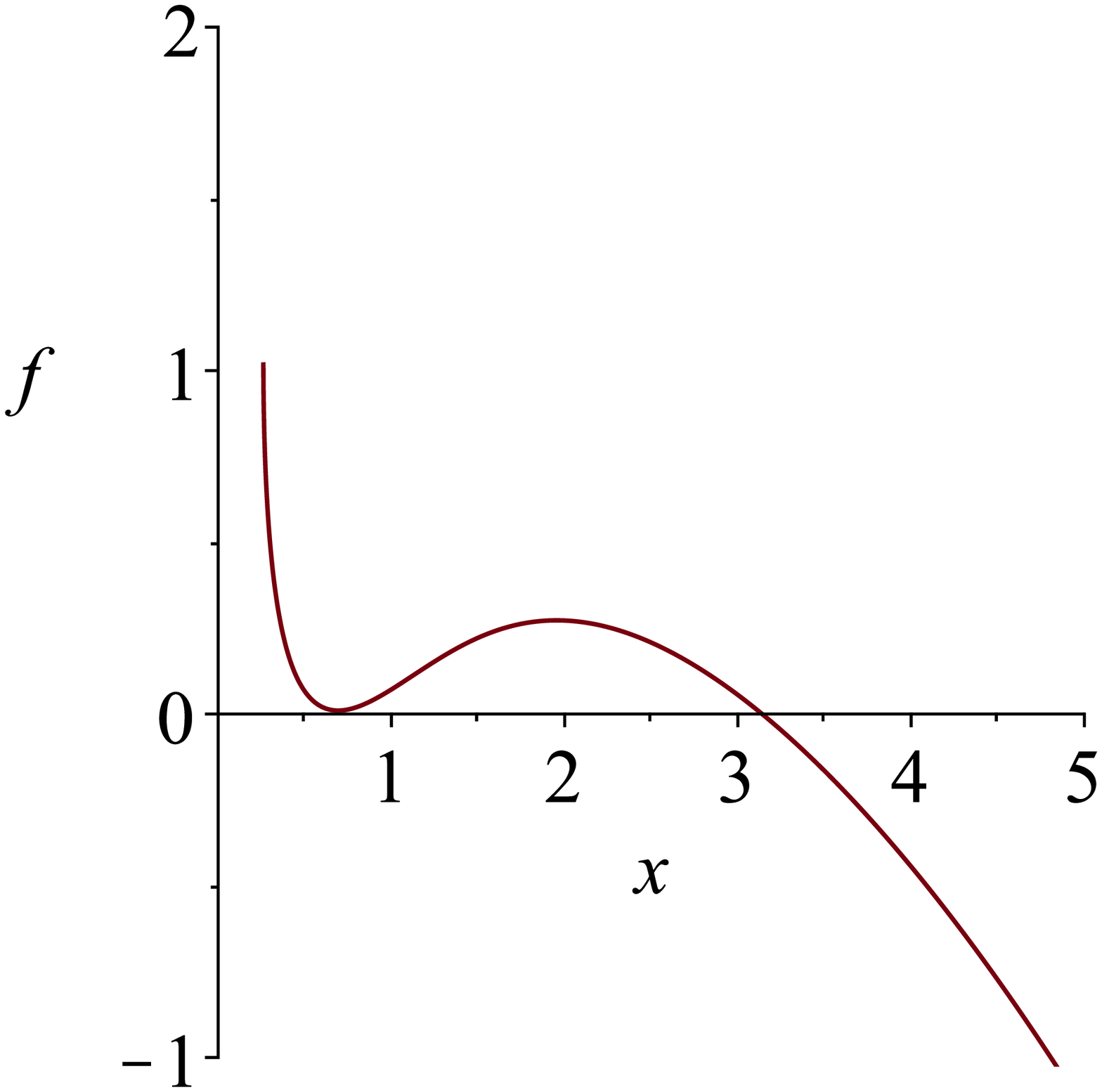}
    \end{subfigure}
    \caption{\textbf{Negative Mass}: $m=-0.3$,\;$z=0.09$,\; $d=6$\ \textbf{Left:}\;$b_{2}=-2$, showing three distinct horizons.  \textbf{Center:}\;$b_{2}=-3$, showing   one double root depicting a Nariai like solution\;\textbf{Right:}\;$b_{2}=-1.3$, showing an extremal black hole with a distinct cosmological horizon.}
    \label{negmass1}
\end{figure}

Negative mass solutions have more interesting behaviour, since if  $b_{2}<0$ there can now be
three roots to \eqref{horloc}. 
We illustrate this in  Figure~\ref{negmass1}. The left image depicts three horizons, while the center and right image show the two possible
cases where horizons can merge. In the center image   $x_{c}=x_{+}$ (a Nariai-type solution)  and in the right $x_{-}=x_{+}$ (an extremal black hole with a cosmological horizon).   These latter two extremes constrain the range of allowed values of $b_2 < 0$ that yield black hole solutions of negative mass. For the case illustrated in
Figure~\ref{negmass1},   we approximately have $-1.3\geq b_{2} >-3$.

We also find that as $x\to 0$, $f$ does not have a smooth limit, and instead
becomes complex at some finite positive value of $x$.  For $m<0$ and $b_{2}<0$ there is \textit{\textbf{only one}} sign change in the coefficients of the terms in \eqref{fbeq} and hence this equation has only one positive root, at which point the Kretschmann scalar \eqref{Kret} diverges. For the admissible range of $b_2$, this singularity is always behind an horizon as can be seen in Figure~\ref{negmass1}. For the more generic 3-horizon case shown in the left diagram in Figure~\ref{negmass1}, 
 the solution to \eqref{fbeq} is $x=0.1999637662$.

\subsubsection{Allowed values of $b_{2}$ for valid negative mass black hole solutions}

To determine the   the range of $b_{2}$ that allows for a valid black hole solution of negative mass in any dimension, we must solve the following set of equations
\begin{align}
    & f=0,\quad \frac{\partial f}{\partial x}=0\  \mt{for}\  b_{2},x \quad \mt{with} \quad \{x,b_{2}|\in \mathbb{R} >0 \}.
\end{align}
The first equation is given by \eqref{horloc} and yields 
\begin{equation}
    b_{2}=x_+^{4} z-x_+^{2}+2 x_+^{5-d} m
\end{equation}
where $x_+$ solves
\begin{equation} \label{xroot}
    2 x_+^{d-1} z-x_+^{d-3}-m(d-5)=0.
\end{equation}
from the latter equation.
There is no general solution to this set of equations except in certain dimensions. 

For $d=6$ equation \eqref{xroot} is a quintic, and there is no analytic solution for arbitrary $m,z$. For the parameters used  in Figure~\ref{negmass1}, we obtain numerically \begin{equation}
    \left\{x_{{1+}}=0.6897087623,\; b_{{2}_{max}}=-1.325264590\right\},\left\{x_{{2+}}=2.328863088,\; b_{{2}_{min}}=-3.033847194\right\}.
\end{equation}
refining the values given above.

In 7 dimensions however, we may find a closed form expression.  The equations become
\begin{align}\label{b2soleq}
        &b_{2}=\frac{x_+^{6} z-x_+^{4}+2 m}{x_+^{2}}\\
        &2 x_+^{6} z-x_+^{4}-2 m=0. \label{soleq}
\end{align}
which admits analytic solutions for arbitrary $m,z$.  The latter equation  is a cubic in $x_+^{2}$; writing $m=-|\textbf{m}|$ we get
\begin{equation}\label{yeq}
    2 y^{3} z-y^{2}+2 |\textbf{m}|=0\quad  \text{with}\quad  y=x_+^2,\ z>0
\end{equation}
Now through Decartes' rule of signs, the possible number of \text{distinct} positive roots is either 2 or zero, while the number of negative roots will \textit{always} be one;
this latter case is inadmissible since we must have $y>0$. 

To have two \textit{distinct} positive real roots for $y$ (and thus the same number of distinct positive values for $x_+$),
we require the discriminant
\begin{equation} \label{discrim}
    \Delta_{3}=8 |\textbf{m}|\left(1-54 z^{2} |\textbf{m}|\right) > 0
\end{equation}
for \eqref{yeq},  implying
\begin{equation} \label{discond}
    -\frac{1}{54 z^{2}}<m<0.
\end{equation}
This condition must be satisfied to have three distinct roots.  Writing
\eqref{yeq} as a depressed cubic, the solutions (for $x$) are easily written as
\begin{equation}
    x_{k}=\sqrt{\frac{1}{3 z} \cos \left(\frac{1}{3} \arccos \left(1-108 \mathbf{|m|} z^{2}\right)-\frac{2 \pi k}{3}\right)-\frac{1}{6 z}} \quad k=0,1,2
\end{equation}
For all values of $m$   satisfying \eqref{discrim}, the only $k = 0,1$
yield real solutions, where $x_{0}>x_{1}$. The value $x_{0}$ corresponds to the location of the location of the Nariai-like black hole solution whereas $x_{1}$ locates the extremal black hole of negative mass. From \eqref{b2soleq}, the respective minimal and maximal values of $b_2$ are 
\begin{equation}
\begin{gathered}
%b_{2}=\frac{z x^{6}-x^{4}-2 \mathbf{|m|}}{x^{2}} \\
b_{{2}_{min}}=\frac{8 \cos (A)^{3}-432 \mathbf{|m|} z^{2}-36 \cos (A)^{2}+30 \cos (A)-7}{72 z^{2} \cos (A)-36 z}   \\
A=\frac{1}{3} \arccos \left(1-108 \mathbf{|m|} z^{2}\right) \\
\\
b_{{2}_{max}}=\frac{8 \cos (B)^{3}-432 \mathbf{|m|} z^{2}-36 \cos (B)^{2}+30 \cos (B)-7}{72 z^{2} \cos (B)-36 z}\\
B=\frac{1}{3} \arccos \left(1-108 \mathbf{|m|} z^{2}\right) -\frac{2 \pi}{3}
\end{gathered}
\end{equation}
and provided $ b_{{2}_{max}} > b_2 > b_{{2}_{min}}$, there will be a valid negative mass  black hole with two horizons and a de Sitter cosmological horizon.

We can make a further  interesting observation   about \eqref{discrim}.
If we saturate the inequality \eqref{discond} from above, setting 
set $ m = 0$, a range of interesting phenomena can occur. It is possible to have a black hole, a Nariai like black hole, pure de Sitter space or a naked singularity. These scenarios are analyzed in the next section.

\section{Massless Black Holes}

The saturation of \eqref{discond} from above allows for the possibility of zero mass exotic black holes. A intriguing feature of these possible black holes is that they are independent of the dimension of the spacetime. Taking \eqref{feq} and setting the mass to zero gives us
\begin{equation}
    f=1+\frac{x^{2}}{2}-\frac{\sqrt{(1+4 z) x^{4}+4-4 b_{2}}}{2}.
\end{equation}
The root equation \eqref{xroot} is now
\begin{equation}
     x^{4} z- x^{2}- b_{2}=0
\end{equation}
which is a biquadratic equation in $x^2$ so it can be written as
\begin{equation}
    y^2 z - y -b_{2}=0.
\end{equation}
We may again analyze the roots of this polynomial by Decartses' rule of signs. Assuming $z>0$, then
\begin{equation}
    \begin{aligned}
    &\text{for}\ b_{2}<0\qquad 2\ \text{or}\ 0\ \text{roots}\\
    &\text{for}\ b_{2}\geq 0\qquad  1\ \text{root}
    \end{aligned}
\end{equation}
As a quadratic the discriminant is immediately written down as: $\Delta_{2}\equiv 4 z b_{2}+1$. For a valid black hole solution, we must have $\Delta_{2}\geq 0$. The saturation of this inequality suggests the Nariai solution. The unsaturated inequality gives us, similar to the negative mass case, a condition on $b_{2}$ as a function of $z$ that must be satisfied for a black hole to exist: $-\frac{1}{4 z}<b_{2}<0$. If $b_{2}$ is not within these bounds there will be a naked singularity located at the origin. 

\begin{figure}[H]
    \centering
    \includegraphics[width=0.19\textwidth]{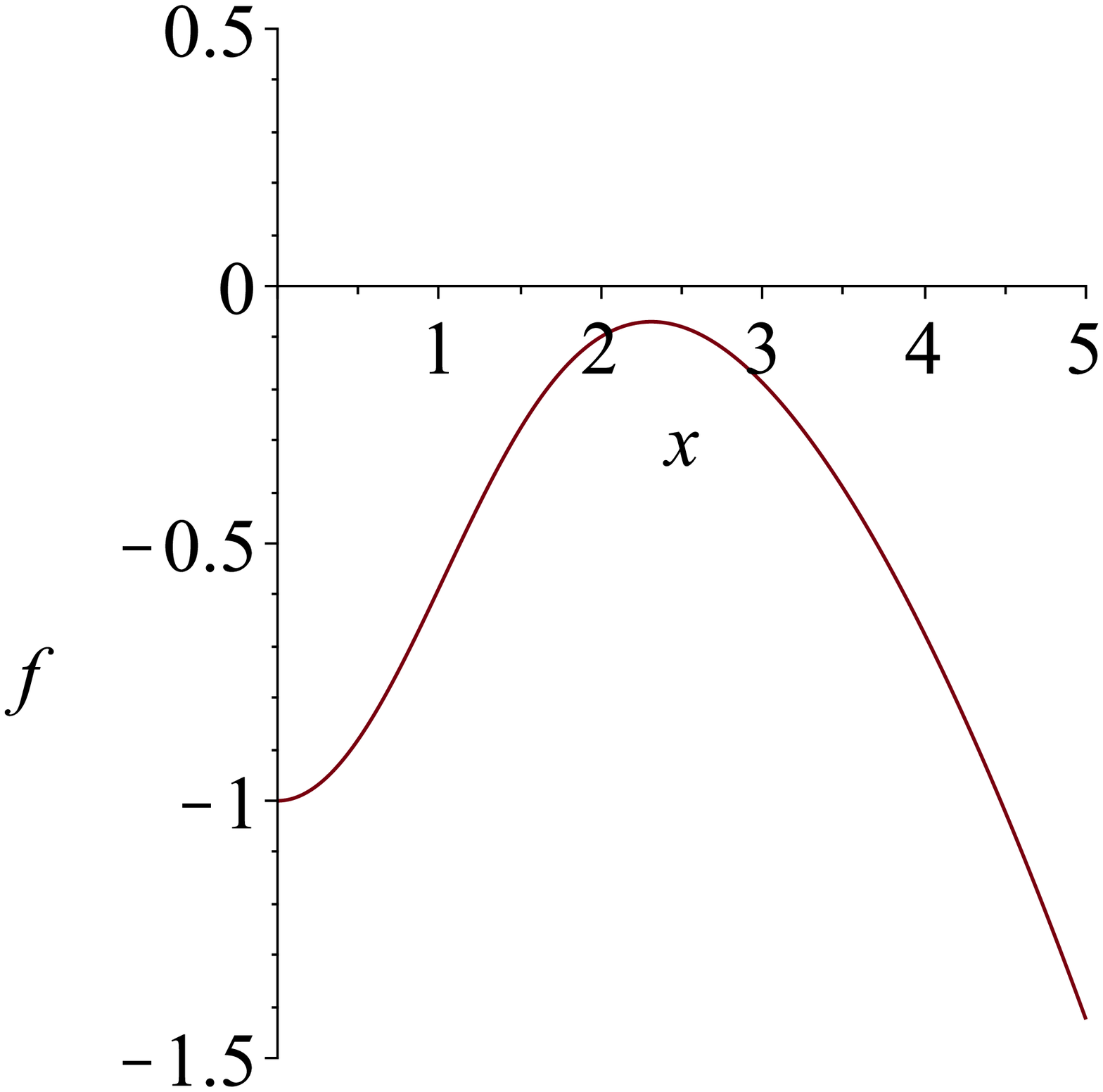}
    \includegraphics[width=0.19\textwidth]{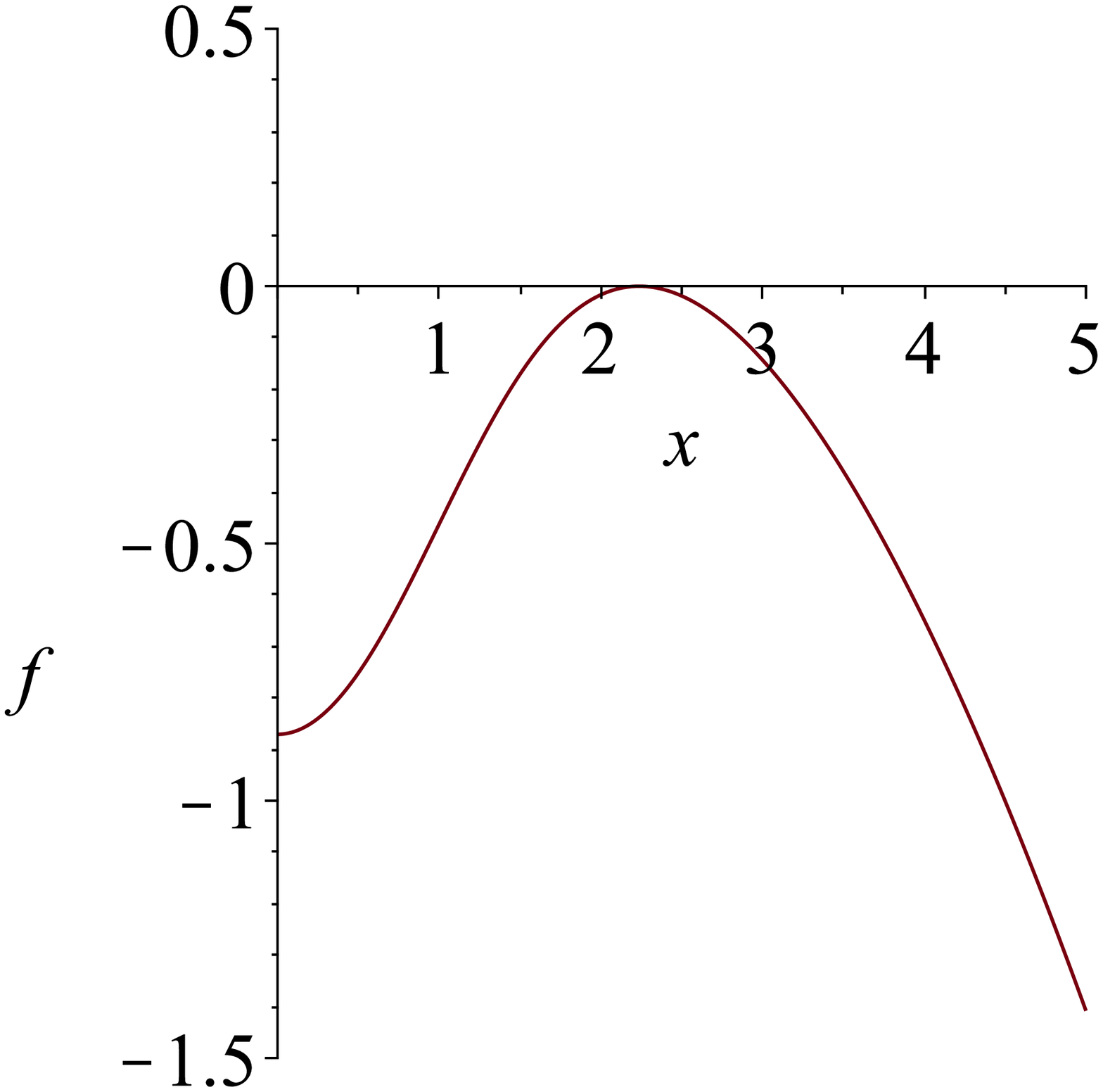}
    \includegraphics[width=0.19\textwidth]{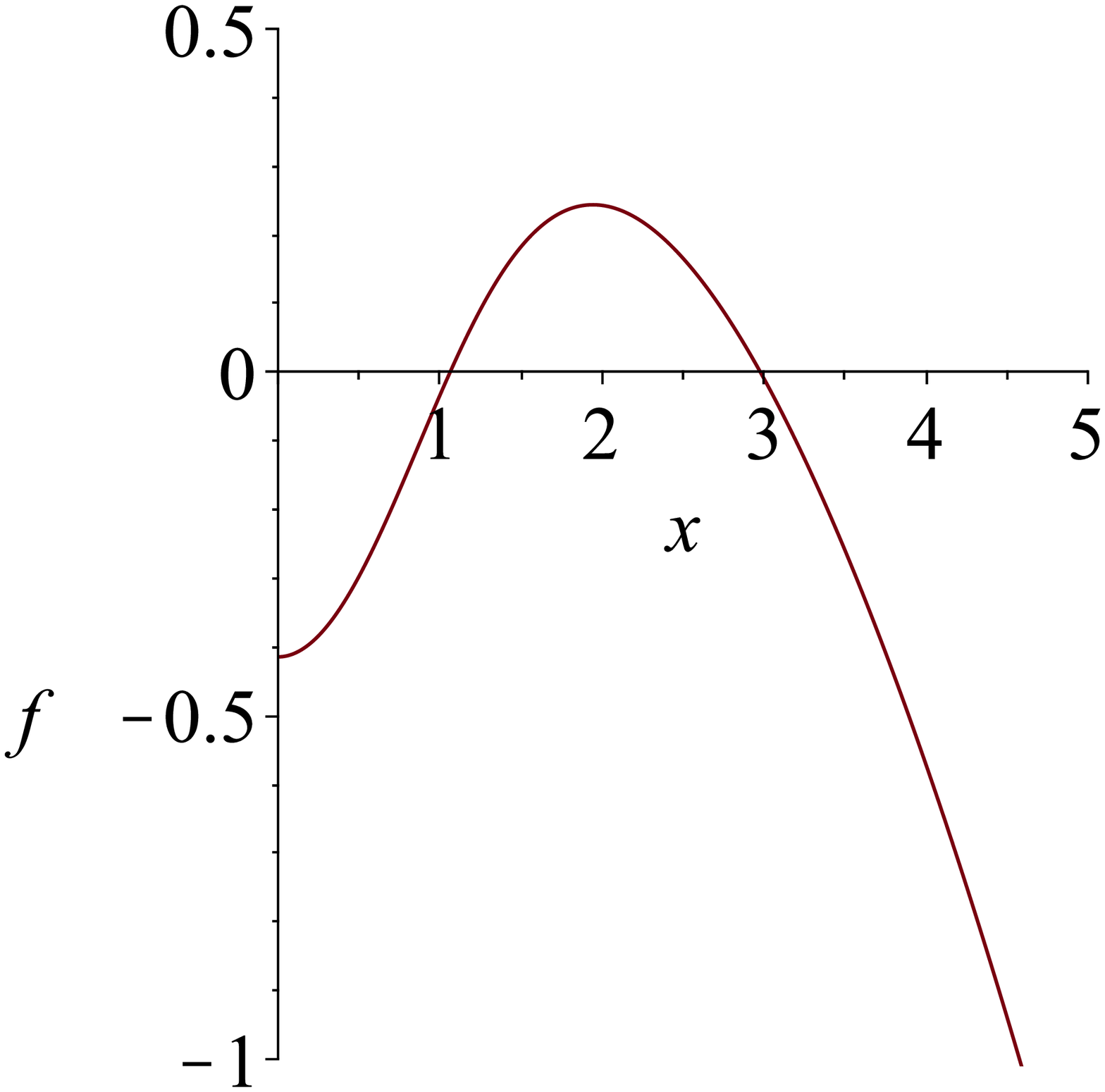}
    \includegraphics[width=0.19\textwidth]{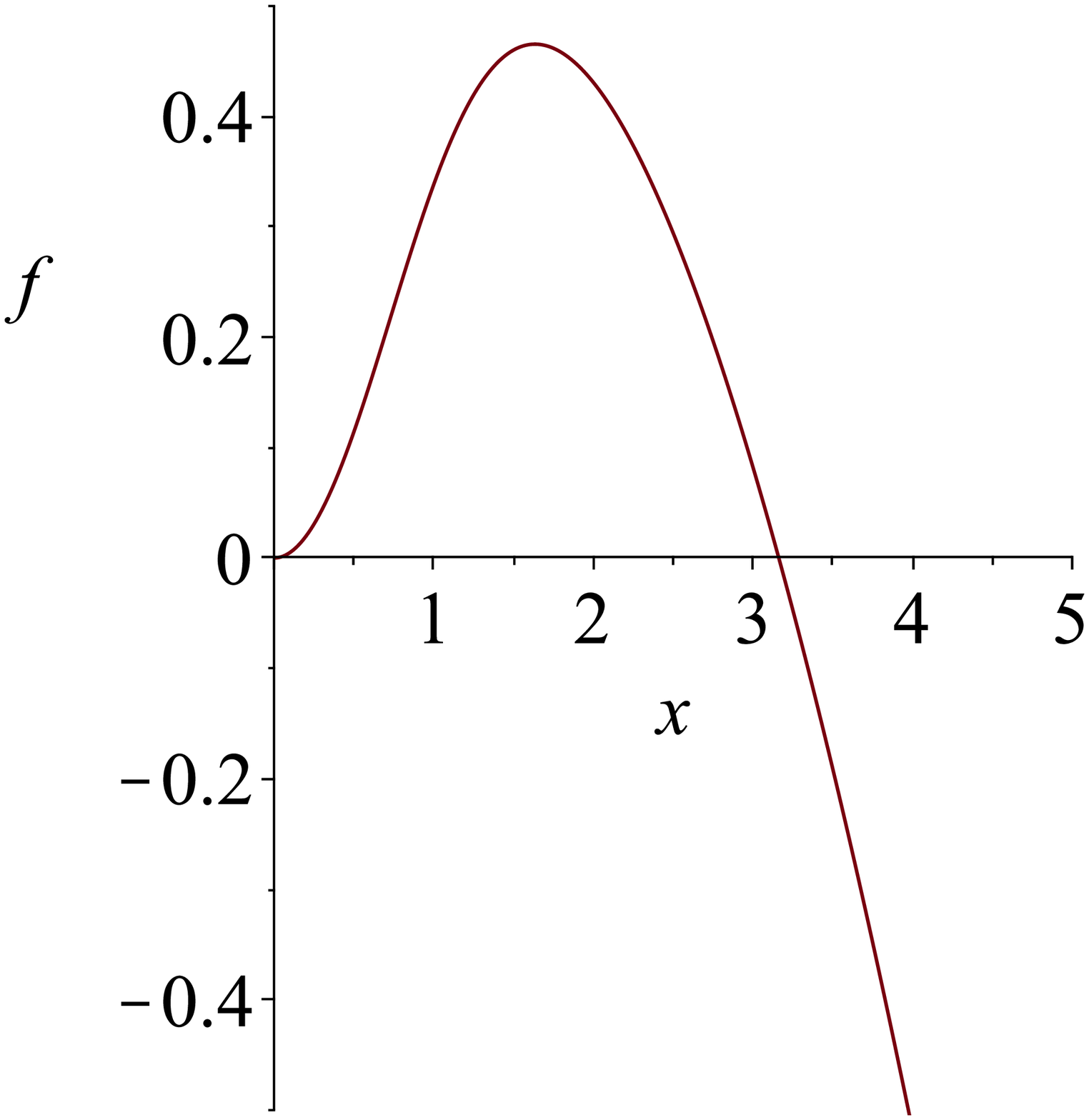}
    \includegraphics[width=0.19\textwidth]{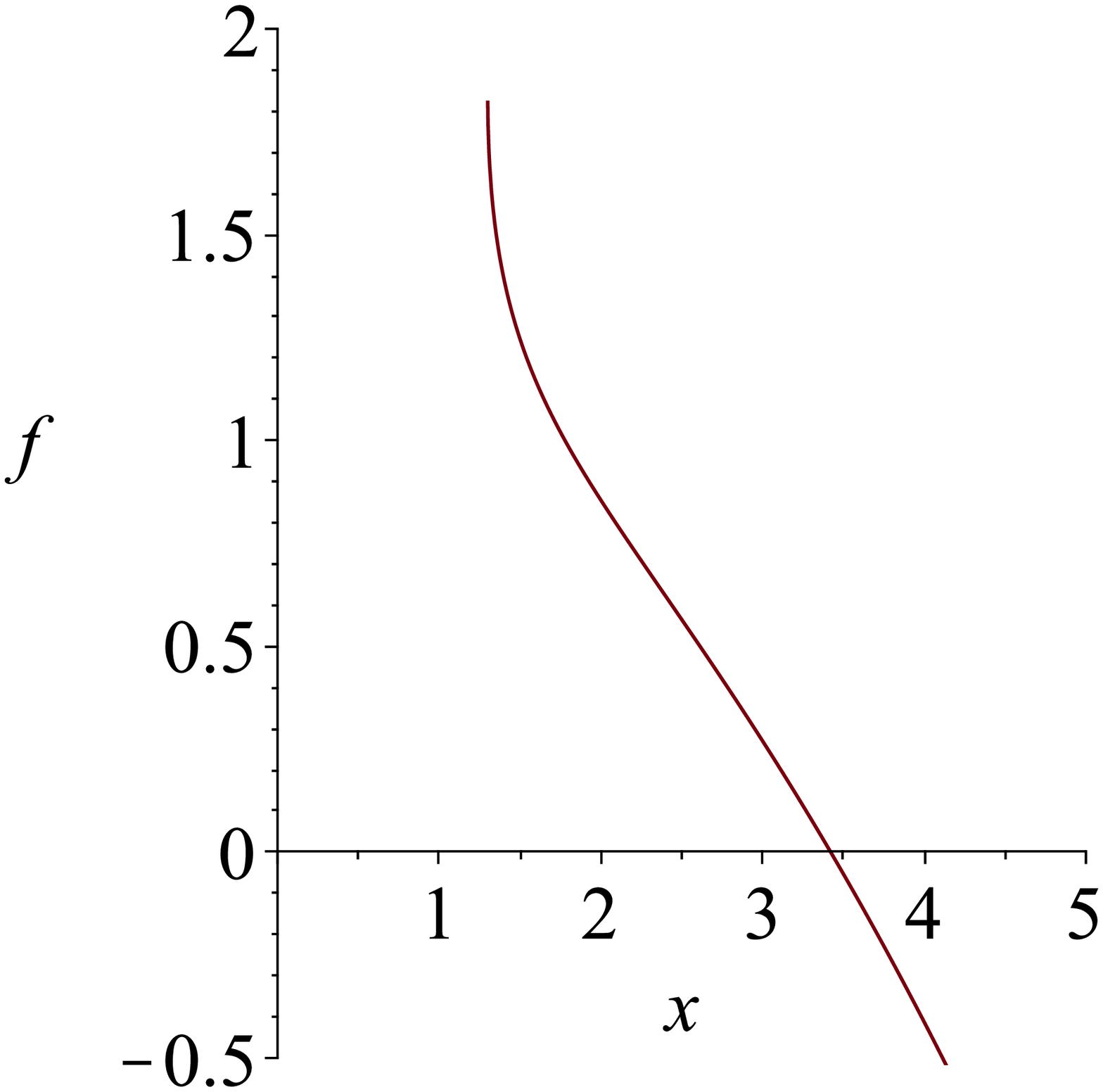}
    \caption{\textbf{Massless Soltuions} $z=0.1$ From left to right the $b_{2}$ values are:-3, -2.5, -1, 0, 2. It depicts a naked singularity at the origin, a Nariai like black hole, a de Sitter black hole, pure de Sitter space, and then a non-trivial naked singularity outside of the origin, respectively.}
    \label{zeromass}
\end{figure}

For $b_{2}>0$ the space is pure de Sitter, provided   \eqref{fbeq} 
has no solutions so as 
to avoid singularities. These equations are easily solved
for any $d$, yielding
\begin{equation} \label{singpos}
 x=\frac{\sqrt{2}\left(b_{2}-1\right)^{1 / 4}}{(1+4 z)^{1 / 4}}.
\end{equation}
and so we conclude that we must have $0\leq b_{2}<1$ to obtain a pure de Sitter space without any naked singularities. The naked singularity that would occur for $b_{2}\geq 1$ is at a finite value of $x$  given  by \eqref{singpos}. The different scenarios are outlined in Figure~\ref{zeromass}. The 3 leftmost images display negative values of $b_{2}$ respectively showing a naked singularity, Nariai black hole and de Sitter black hole. The two right most images are for zero and positive values of $b_{2}$ respectively, with the former being pure de Sitter space and the latter a naked singularity outside of the origin. For increasing values of $b_{2}>0$ the termination point of $f$ at $x=0$ moves up the vertical axes until it reaches a maximum point at $f(x,b_{2},z)\rightarrow f(0,1,0.1)=1$.

\section{Conclusion}

We have shown that metrics of the form \eqref{metric}, with $\gamma_{\alpha \beta}$ being the metric of an arbitrary Einstein space,  can possess both negative mass and massless 
asymptotically de Sitter black hole solutions in Gauss-Bonnet  gravity. These $M\leq 0$ black holes are all exotic black holes, and their existence  depends on the relative values of the parameters $b_{2}$, $z$ and $m$. For the massless case, for there is a direct relationship between $b_{2}$ and $z$ with $b_{2}$ being bounded from below by $z$. 
For $M<0$ black holes there are more restrictions on the parameters. The range of $m$ is bounded by $z$,
as given by \eqref{discond}.  There is also  a minimum and maximum allowed value 
of $b_2$ admitting such solutions.
 
We close by noting that the existence of such horizon geometries is still an open question. This is due to the fact that field equations for the transverse space are an over constrained PDE system, where certain topological parameters $b_{n}$ may not provide a solution. However we are not aware of any theorem forbidding the existence of non-trivial solutions to  \eqref{exoticeqs}.

With the discovery of negative mass black holes in de Sitter space interesting questions arise. These include the behaviour of geodesics, formation from gravitational collapse, and possible pair production in the early universe of such objects.   Their thermodynamic behaviour is also likely to yield interesting surprises in comparison to their Anti de Sitter counterparts 
\cite{Hull2021,HullSimovic}.

\section*{Acknowledgements}
This work was supported in part by the Natural Sciences and Engineering Research Council of Canada.

\newpage
\bibliographystyle{unsrt}
\addcontentsline{toc}{section}{References}
\bibliography{BIB.bib}

\end{document}